\documentclass[%
 reprint,
 amsmath,amssymb,
 aps, prl, 
longbibliography
]{revtex4-2}

\usepackage{graphicx}
\usepackage{dcolumn,color}
\usepackage{bm,comment}
\DeclareMathOperator{\sech}{sech}
\DeclareMathOperator{\cosech}{cosech}

\let\originalvec\vec
\renewcommand\vec[1]{\originalvec{\kern0pt #1}}

\newcommand\bx{ {\bf x}}
\begin{document}

\preprint{APS/123-QED}

\title{Mean first passage times of higher-dimensional velocity jump processes}

\author{Maria R. D'Orsogna}
\email{dorsogna@csun.edu}
 \affiliation{Department of Mathematics,
 California State University at
 Northridge,  Los Angeles, CA, 91330, USA}
 
 \affiliation{Department of Computational Medicine, 
 University of California at Los Angeles, Los Angeles, 90095-1766, CA, USA}
\author{Alan E. Lindsay}%
\affiliation{Department of Applied and Computational Mathematics and Statistics, University of Notre Dame, Notre Dame, IN, 46656,
USA
}

\author{Thomas Hillen}
\affiliation{
Department of Mathematical and Statistical Sciences, University of Alberta, Edmonton, T6G 2G1, Canada
}%

\date{\today}

\begin{abstract}
First passage phenomena arise across physics, biology, and finance when stochastic processes first reach a threshold, triggering downstream events. Examples include the irreversible exit from a domain, a biochemical reaction, a financial selloff. While typical formulations involve diffusive motion, many stochastic processes are better described as velocity jump processes, characterized by persistent motion interrupted by stochastic velocity changes. Despite their ubiquity, first-passage properties of velocity jump processes remain underdeveloped in higher dimensions, especially under directional bias. We present a general framework to estimate the mean first passage time (MFPT) and higher moments of the survival probability for fixed-speed velocity jump processes where possible reorientations range from strong alignment to full angular anisotropy. For low Knudsen numbers, when the mean free path is small compared to the distance to the target, we derive a universal form for the MFPT in which two bias functions encode broad classes of angular distributions, including von Mises–Fisher, wrapped Cauchy, and elliptical families. In the narrow capture limit of a vanishingly small target, directional persistence induces anomalous scaling, including regimes where the MFPT remains finite whereas standard diffusion would predict divergence. Finally, we obtain a Langevin representation that accurately reproduces first-passage statistics. Analytical predictions are confirmed
by numerical simulations.
\end{abstract}

\maketitle

Velocity jump processes capture a wide range of realistic non-diffusive 
behaviors \cite{VanKampen1981, Klages2008}. 
Examples include molecular collisions in biochemistry \cite{Hanggi1990},
run-and-tumble motion in bacteria \cite{Newby2011,Scacchi2018, Dhar2019, Angelani2014,iyaniwura2025meanpassagetimeactive}, 
pricing in quantitative finance \cite{Bielecki2004, Masoliver2008, Perello2011}, 
animal movement \cite{Fauchald2003, Benichou2011, Kurella2015, Costa2024}.
In all cases, the items of interest ``run" along a given direction for a random time
until a new velocity is selected.
The resulting dynamics is intermittent directed motion that can significantly differ from standard
diffusion due to the inherent persistence. 
Velocity jumps may arise from internal mechanisms, 
external perturbations, or in response to sensory cues. 
The best known 
representation of this paradigm
are the one-dimensional telegrapher equations
that are commonly used to study run-and-tumble behaviors \cite{Heaviside1970, hillen2013transport,HillenReview,Angelani2015}. 
Non-equilibrium first-passage properties of jump processes
are of great significance and may be even more important than 
steady-state quantities such as
moments of the mean displacement or particle distributions \cite{Datta2024}. 
For example, the mean first passage time (MFPT), defined as 
the average time to first reach a target  \cite{Redner2001}, informs 
how long one must wait for downstream, often irreversible, events to
be initiated, such as immune responses, biochemical transformations,  
entry into subcellular compartments \cite{Ghusinga2017, Dorsogna2009, Nowak2009}, 
escape from an enclosed domain \cite{Condamin2008, Bressloff2013, Gardiner2004}. 
The time to reach a threshold is also used as a measure
of resilience in ecology \cite{Arani2021} or to quantify the onset of 
mental health disorders \cite{Mao2024}. 
Understanding how persistence and directionality 
affect 
first passage phenomena is critical  \cite{
Camley2014, Sevilla2019, Artuso2018, Klinger2023, Vezzani2024}; 
however, the problem is notoriously difficult in 
higher dimensions because velocities can reorient
along a continuum of directions, and 
not just to the left or right
as in $d=1$  \cite{Rupprecht2016, Grebenkov2016, Mori2020}.

The aim of this Letter is to help fill this gap. 
We write a general equation for the MFPT $\Theta({\bf x},{\bf {v}})$ 
to a given target for a particle starting at position ${\bf {x}}$ with velocity ${\bf {v}}$ under a velocity jump process 
where jumps occur at rate $\mu$ and where $q({\bf {x}},{\bf {v}})$ is the memoryless probability 
density function for a particle at  ${\bf {x}}$ to switch to velocity ${\bf {v}}$.
In many settings the speed $v$ is fixed,  
to a physiological value in biology or to the thermal velocity in Brownian motion for example, 
so that the jump process becomes a series
of stochastic reorientations.  The turning angle may be completely random, 
as in run-and-tumble systems, 
or may bias the motion along a preferred direction, as when there is an underlying 
network, external field or chemical trail. For fixed speed
$ v  = \sigma$ and
small Knudsen number $\varepsilon = \sigma / (\mu L)  \ll 1$, 
where $\sigma / \mu$ is the mean free path and $L$ is a characteristic distance 
the particle must travel before reaching the target or escaping the domain,  
we show that $\Theta({{\bf {x}}},{\bf {v}}) \sim T({\bf {x}}) + {\cal O}(\varepsilon)$
and write a self-contained differential equation for $T({\bf {x}})$ 
valid in $d$-dimensional systems.
We solve for $T({\bf {x}})$ under
representative turning angle distributions in simple geometries
and compare our results with numerical simulations for $d=2, 3$.

The probability density $P ({\bf {x}}, {\bf {v}}, t)$
to be at position ${\bf {x}}$ with velocity ${\bf {v}}$ at time $t$ obeys the forward equation
\begin{equation}
\label{forward}
\frac{\partial P }{\partial t}  + {\bf {v}} \cdot  \nabla P = - \mu P +  \mu \int q({\bf {x}},{\bf {v}}) P({\bf {x}}, {\bf{v}'}, t) d \bf{v}'
\end{equation}
for given initial and boundary conditions.
The integral is over the chosen velocity domain
of $q({\bf {x}}, {\bf {v}})$. 
The survival probability 
$S({\bf {x}}, {\bf {v}},t)$ for a particle initiated 
at $({\bf {x}}, {\bf{v}})$ to remain in a spatial domain
$\cal D$ without having reached a specified target at time $t$ follows the backward 
Kolmogorov equation, given 
by the adjoint of Eq.\,\ref{forward}
\begin{equation}
\label{backward}
\frac{\partial S }{\partial t}  - {\bf {v}} \cdot  \nabla S = - \mu S +  \mu \int q({\bf {x}},{\bf{v}}') S({\bf {x}}, {\bf{v}'}, t) d \bf{v}', 
\end{equation}
where $S({\bf {x}}, {\bf {v}}, 0) =1$ for all ${\bf {x}}$ in $\cal{D}$ and any $\bf v$. 
Eq.\,\ref{backward} is solved with
absorbing boundaries at the
target and reflecting elsewhere; more details are 
given in the Supplemental Material (SM). 
The escape probability $1 - S$ yields the first-passage time probability density 
 $-{\partial S}/ {\partial t}$. The MFPT $\Theta({\bf {x}}, {\bf {v}})$ is its first moment
\begin{equation}
\label{moment}
\Theta({\bf {x}}, {\bf {v}}) = - \int_{0}^{\infty} t \frac{\partial S} {\partial t} dt =  \int_{0}^{\infty} S( {\bf {x}}, {\bf {v}}, t) dt. 
\end{equation}
Integrating Eq.\,\ref{backward} over time yields
\begin{equation}
\label{MFPTgen}
-1  - {\bf {v}} \cdot  \nabla \Theta = - \mu \Theta +  \mu \int q({\bf {x}},{\bf {v}}') \Theta({\bf {x}}, {\bf {v}}') d {\bf {v}}'.
\end{equation}
Upon multiplying Eq.\,\ref{MFPTgen} by $q({\bf {x}}, {\bf {v}})$ and 
by $\bf {v} \, q({\bf {x}}, {\bf {v}})$,  integrating both expressions 
over velocity space, and merging results one obtains 
\begin{eqnarray}
\nonumber
-1 &=& \frac{1}{\mu} \int q({\bf {x}}, {\bf {v}}) ({\bf {v}}  \, \otimes \, {\bf {v}} : 
\nabla \, \otimes \, \nabla) \Theta({\bf {x}}, {\bf {v}})  \, d {\bf {v}} + \\
\label{MFPT}
&&  {\bf {b}}( {\bf {x}})\cdot {\nabla} \int q({\bf {x}}, {\bf {v}}) \,\Theta ({\bf {x}}, {\bf {v}}) d {\bf {v}}. 
\end{eqnarray}
Here $\otimes$ represents the tensor product
between vectors and the colon their convolution, so that
in dimension $d$
\begin{eqnarray}
\nonumber
{\bf {v}}  \, \otimes \, {\bf {v}} : 
\nabla \, \otimes \, \nabla =  \sum_{i,j = \{1,2, \dots, d \}} v_i v_j \frac{\partial}{\partial{x_i}} \frac{\partial}{\partial{x_j}}, 
\end{eqnarray}
and ${\bf {b}} ({\bf {x}})$ is a drift term given by 
\begin{eqnarray}
\label{drift}
{\bf {b}} ({\bf {x}}) = \int {\bf {v}} \, q({\bf {x}}, {\bf {v}}) d {\bf {v}}. 
\end{eqnarray}
If $q({\bf {x}}, {\bf {v}}) = q({\bf {x}}, -{\bf {v}})$
reorientation is symmetric in ${\bf v}$ and ${\bf {b}} ({\bf {x}}) =0$ \cite{Hillen2025}.
We now fix the speed $v = \sigma$. For $d=1$ 
Eqs.\,\ref{forward} and \ref{backward} yield the
 telegrapher's equations with $q(x, \sigma) = q(x,-\sigma) = 1/2$
\cite{Weiss2002, Weiss1994, Hemmer1961, Kac1974, DOrsogna2003}.  
For $d >1 $,
we set ${\bf v} = \sigma \, \hat \theta$, where
$\hat \theta $ is the unit vector along the direction of motion. We write
$q({\bf {x}}, {\bf {v}}) = q({\bf {x}}, \hat \theta) \, \delta(v - \sigma) / \sigma^{d-1} $ 
where $q({\bf {x}}, \hat \theta)$ is the reorientation distribution and
denote the angular volume element $d \hat \theta_{d-1}$ so that
$\int q({\bf {x}}, \hat \theta) d \hat \theta_{d-1} = \int q({\bf {x}}, {\bf {v}}) d {\bf {v}} = 1$ and
${\bf {b}}({\bf {x}}) = \sigma \int \hat \theta q({{\bf {x}}}, \hat \theta) d \hat \theta_{d-1}$.
Using the non-dimensional variables ${\bm \xi} = {\bf {x}}/L$,  
$\tau = \sigma^2 t / (\mu L^2)$, with $L$ a typical length scale, we rewrite
Eq.\,\ref{backward} as
\begin{figure}[t!]
\includegraphics[width=0.35\textwidth]{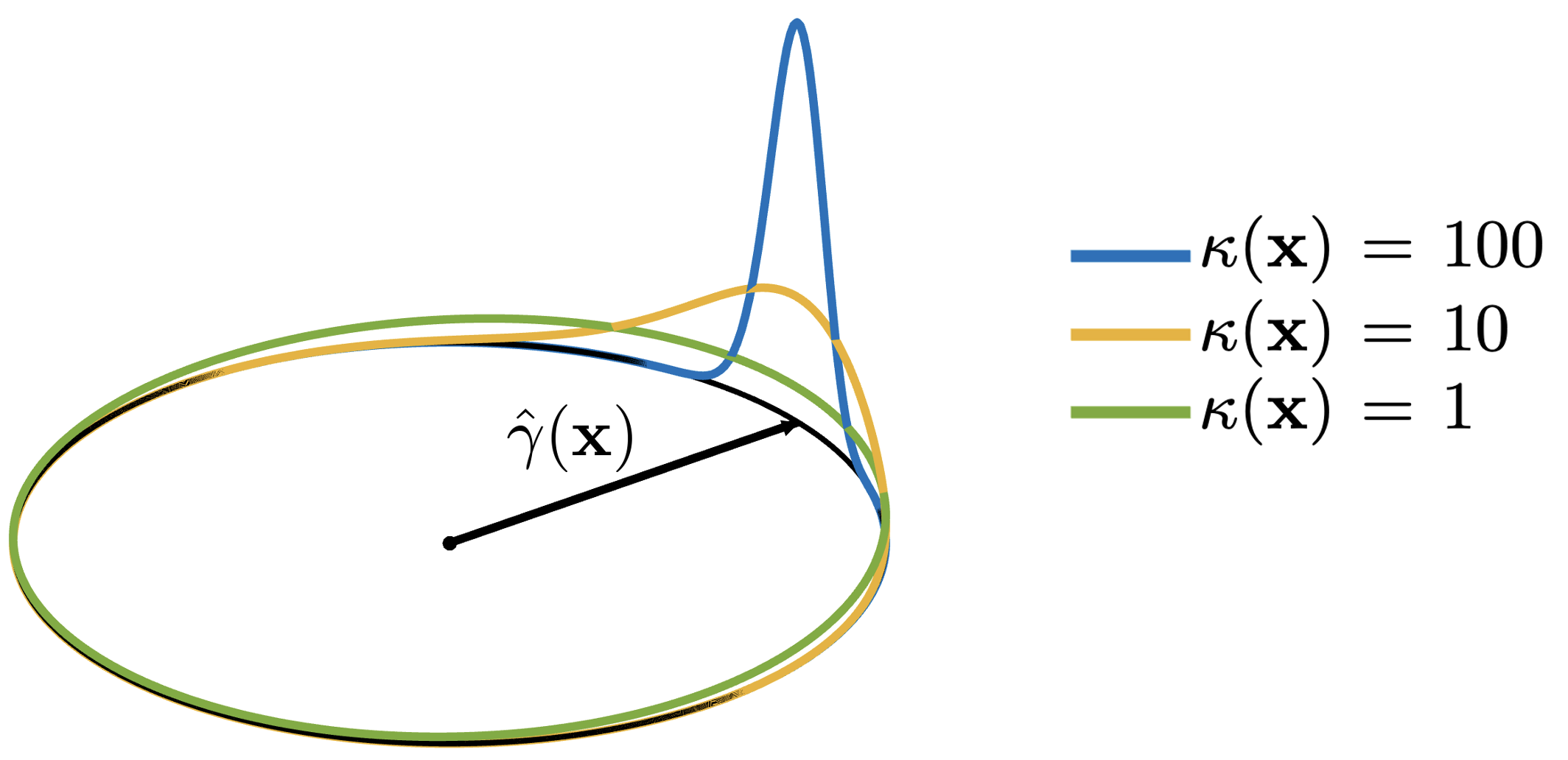}
\caption{
The von Mises angular distribution in Eq.\,\ref{VM} for $d=2$ and three 
values of the concentration $\kappa(\bf x)$ that quantifies the
bias towards $\hat \gamma(\bx)$. 
Larger $\kappa (\bf x)$ yields sharper alignment.}
\end{figure}
\begin{equation}
\label{backwardres}
\varepsilon^2 \frac{\partial S }{\partial \tau}  - \varepsilon \, \hat \theta \cdot  \nabla_{{\bm {\xi}}} S = - S +   \int q({\bm {\xi}}, \hat \theta) S({\bm {\xi}}, \hat \theta, \tau) d \hat \theta_{d-1}.
\end{equation}
Here, $\varepsilon = \ell/ L$ is the Knudsen number and $\ell = \sigma/\mu$ the mean free path. We now assume $\varepsilon \ll 1$. 
Due to the drift term, a standard asymptotic expansion is insufficient,  so we separate fast and slow time scales
by introducing $z = \tau/ \varepsilon$ and pose
$S({\bm {\xi}}, \hat \theta,\tau) = S_0({\bm {\xi}}, \hat \theta, \tau) + 
\varepsilon S_1({\bm {\xi}}, \hat \theta, \tau)  + \dots$  where 
 $S_i({\bm {\xi}}, \hat \theta,\tau) = s_i({\bm {\xi}}, \hat \theta, \tau, z)$ \cite{Bender1999}. 
The initial condition $S({\bf {x}}, {\bf {v}}, 0) =1$ for all ${\bf {x}}$ in $\cal D$ 
for any $\bf v$ maps to $s_i({\bm {\xi}}, \hat \theta, 0,0)= 0$ for $i \neq 0$ and
 $s_0({\bm {\xi}}, \hat \theta, 0,0)= 1$. Thus, 
\begin{eqnarray}
\label{derivativeres}
\frac{\partial S} {\partial \tau} = \frac 1 {\varepsilon} \frac{\partial s_0}  {\partial z} + 
 \left( \frac{\partial s_0}{\partial \tau } + \frac{\partial s_1} {\partial z} \right)+ 
\dots 
 \end{eqnarray}
By inserting Eq.\,\ref{derivativeres} into Eq.\,\ref{backwardres}, 
matching orders in $\varepsilon$ and returning to $\tau$,  
we find that $S_0({\bm {\xi}}, \tau)$  is independent
of $\hat \theta$ and that the integral $\int_0^{\infty} S_0 ({\bm {\xi}}, \tau) d\tau$ satisfies a self-consistent differential equation. 
This integral is the rescaled MFPT to leading order.  Rewriting it in dimensional units, $T({\bf {x}}) = (\mu L^2 /\sigma^2) \int_0^{\infty} S_0 ({\bf {x}}, \tau) d\tau $, yields a self-consistent differential equation for $T({\bf x})$, 
valid for $\varepsilon \ll 1$, given by
\begin{eqnarray}
\label{superfinal}
{\mathbb D}({\bf {x}}) : \nabla \otimes \nabla \,
T({\bf {x}}) 
+ {\bf {b}}({\bf {x}}) \cdot \nabla \, T({\bf {x}}) = -1, 
\end{eqnarray}
where the diffusion tensor ${\mathbb D}({\bf {x}})$ is defined as
\begin{eqnarray}
\label{difftensor}
{\mathbb D}({\bf {x}}) = \frac{\sigma^2}{\mu} \int  q ({\bf {x}}, \hat \theta) \hat \theta \otimes \hat \theta \,  d \hat \theta_{d-1}, 
\end{eqnarray}
and $T({\bf x}) =0$ at the target. Eq.\,\ref{superfinal} is one of our main findings and coincides with Eq.\,\ref{MFPT} for $v = \sigma$ and $\Theta({\bf {x}}, {\bf v})= T({\bf x})$. 
It is more general than previous results as it allows directional asymmetries and a non-zero drift. 
More details are in the SM. Locally, interpreting $L$ as the distance between the starting point and the target, 
Eq.\,\ref{superfinal} 
implies that the MFPT is independent of the initial orientation
if the distance to the target is much larger than the mean free path.  
In this case, the many particle reorientations erase memory of the initial heading. The full kinetic description in 
Eq.\,\ref{MFPT} must be retained if instead
the initial position is at ${\cal O}(\ell)$ from the target. 
\begin{table*}[t]
\renewcommand{\arraystretch}{2.3}
\caption{\label{table1}
Bias terms $\alpha({\bf x}), \beta(\bf x)$ for representative reorientation distributions $q({\bf x}, \cdot)$ that direct motion towards
    $\hat \gamma(\bf x)$ through the concentration $\kappa(\bf{x})$ \cite{Wu2000}. 
    For $d=2$, $q({\bf x}, \cdot) = q({\bf x}, \hat \theta)$ = $q({\bf x}, \theta)$; for $d=3$, $q({\bf x}, \cdot)$ = $q({\bf x}, \hat n)$, 
    where, respectively,  $\hat \theta =(\cos \theta, \sin \theta)$ and
    $\hat n = (\sin \theta \cos \phi,  \sin \theta \sin \phi, \cos \theta)$. 
    For set boundary conditions, the MFPT is approximated by Eqs.\,\ref{superfinal}, \ref{diffVM}, \ref{driftVM}, 
    $D = \sigma^2 / (d \mu)$.}
\begin{ruledtabular}
\begin{tabular}{llll}
 distribution &  $q({\bf {x}}, \cdot)$ & $\alpha({\bf {x}})$  & $\beta({\bf {x}})$  
 \\[1ex] 
 \hline
       von Mises (2D) & $q_{\rm M}({\bf {x}}, \theta) = \displaystyle{\frac{e^{\kappa ({{\bf {x}}}) \cos (\theta - \gamma({\bf {x}}))}}{2 \pi I_0(\kappa ({\bf {x}}))}}$  
       & $ \displaystyle{\frac{I_2(\kappa ({\bf {x}}))}
{I_0(\kappa ({\bf {x}}))}} $ & $ \displaystyle{ \frac{I_1(\kappa ({\bf {x}}))}
{I_0(\kappa ({\bf {x}}))}}$ 
  \\[1ex]
\hline
 Fisher (3D) & 
$q_{\rm F}({\bf {x}}, \hat{n}) = \kappa({\bf {x}}) \displaystyle{\frac
{e^{\kappa ({{\bf {x}}}) \hat \gamma({\bf {x}}) \cdot \hat n}}
{4 \pi \sinh(\kappa ({\bf {x}}))}}$  & $ \, \displaystyle{ 1 - \frac 3 {\kappa({\bf {x}})} \coth(\kappa({\bf {x}}))+  \frac 3 {\kappa^2({\bf {x}})} \, }
 $ & $ \, \displaystyle{\coth(\kappa({\bf {x}})) -  \frac 1 {\kappa({\bf {x}})} } \,$ 
  \\[1ex]
\hline
wrapped Cauchy (2D) & $q_{\rm C}({\bf {x}}, \theta) = \displaystyle{\frac{\sech
(2  \kappa({\bf x}))}
{2 \pi (1 - \tanh(2 \kappa({\bf {x}}))  \cos(\theta - \gamma({\bf {x}}))) } }$ &  
 $\displaystyle{\tanh^2(\kappa({\bf {x}}))}$ & $\tanh(\kappa(\bf {x}))$
 \\[1ex]
\hline
Elliptical (2D) & $q_{\rm E}( {\bf {x}}, \theta) = 
\displaystyle{\frac{\sech^{3} 
( \kappa({\bf x}))
}
{2 \pi (1 - \tanh(\kappa({\bf x}))  \cos(\theta - \gamma({\bf x})))^2}} \quad$
& $1 - 2 \cosech^2  (\kappa({\bf x})) (1 - \sech (\kappa(\bf x)))$ \quad
 & $ \tanh (\kappa({\bf x}))$ 
 \\[1ex]
\end{tabular}
\end{ruledtabular}
\label{distributions}
\end{table*}
\noindent
For isotropic reorientations, $q({\bf {x}},\hat \theta) = 1/ S_{\rm d}$, where
$S_{\rm d} = 2 \pi^{d/2}/ \Gamma (d/2)$ 
is the surface area of the unit $d-1$ sphere.  
A direct calculation yields
 ${\bf {b}}({{\bf {x}}}) =0$ and  
${\mathbb D}({\bf {x}}) = D \, {\mathbb I} $,  where $D = \sigma^2 / (d \mu) $
and ${\mathbb I} $ are, respectively, 
the diffusion constant and identity matrix. 
Thus, Eq.\,\ref{superfinal} reduces to the standard diffusive MFPT 
equation $D \, \nabla^2 T({\bf {x}}) = -1$. 
We now evaluate $T({\bf {x}})$ in $d=2$ for
the von Mises distribution, 
the circular analog of 
a Gaussian, that biases reorientations towards
a preferred direction
\cite{Bica2017, 
Hillen2017, SwanHillen}
\begin{eqnarray}
\label{VM}
q_{\rm M}({\bf x}, \hat \theta) = q_{\rm M} ({\bf {x}},  \theta) = \frac{e^{\kappa ({{\bf {x}}}) \cos (\theta - \gamma({\bf {x}}))}}{2 \pi I_0(\kappa ({\bf {x}}))}. 
\end{eqnarray}
\noindent 
Here, $\hat \theta = (\cos \theta, \sin \theta)$, $I_i(\cdot)$ is the modified Bessel function of the first kind of order $i$,  
$\hat {\gamma}({\bf {x}}) = (\cos \gamma ({\bf {x}}), \sin \gamma ({\bf {x}}))$ is the preferred direction at ${\bf {x}}$, 
and $\kappa({\bf {x}}) \geq 0 $ is a unitless concentration parameter, which quantifies
the degree of bias towards $\hat {\gamma}({\bf {x}})$. 
The diffusion tensor and drift are
%
\begin{subequations}
\begin{eqnarray}
\label{diffVM}
{\mathbb D} ({\bf {x}}) &=& D 
\left( 1 - \alpha({\bf {x}})
\right) { \mathbb I}
+ 2 D \alpha ({\bf {x}}) \hat {\gamma}({\bf {x}}) \otimes \hat{\gamma} ({\bf {x}}) \\[4pt]
\label{driftVM}
{\bf b}({\bf {x}}) &=& \sigma \beta({\bf {x}})  \hat \gamma({\bf {x}})
\end{eqnarray}
\end{subequations}
\noindent
where the spatial dependence of $\alpha({\bf {x}})$ and $\beta({\bf {x}})$
is through $\kappa ({\bf {x}})$ as shown in Table \ref{distributions}. 
In the weak-bias limit $\kappa \to 0$, $q_{\rm M}  ({\bf {x}},  \theta)$ becomes uniform; 
in the strong-bias limit $\kappa \to \infty$, $q_{\rm M}  ({\bf {x}},  \theta)$ approaches a Dirac-delta 
along $\hat \gamma({\bf {x}})$.  Further discussions of Eq.\,\ref{VM} and its
limits are in the SM. Since $0 \leq \alpha(\kappa), \beta(\kappa) < 1$ are monotonically increasing in $\kappa$, 
they can be used as indicators of 
 the strength of the bias along $\hat \gamma ({\bf {x}})$;
when both approach zero, the diffusive MFPT equation is recovered;
when both approach unity, motion is perfectly ballistic along $\hat \gamma ({\bf {x}})$.
Remarkably, the same structure shown in Eqs.\,\ref{diffVM} and \,\ref{driftVM}
emerges for other directed distributions in $d=2$ and 
for the Fisher distribution, the analogue of the von Mises distribution in $d=3$, provided all corresponding quantities
are interpreted accordingly. 
We list these distributions in Table \ref{table1} with the corresponding bias terms $\alpha({\bf {x}}), \beta({\bf {x}})$. 
For all of them, $\kappa(\bf x)$ interpolates between the uniform ($\kappa({\bf x}) \to 0$)
and Dirac-delta ($\kappa({\bf x}) \to \infty$) distributions. 
While Eqs.\,\ref{superfinal}, \ref{diffVM}, \ref{driftVM} are 
valid for any preferred direction,  
simple geometries and specific choices for $ \hat{\gamma} ({\bf {x}})$
allow for simplifications. Typical ecological or biological 
scenarios involve finding the MFPT $T({\bf {x}})$
to the boundary of a circular ($d=2$) or spherical ($d=3$) domain 
of radius $R_0$ for preferentially radial motion
so that $\hat \gamma({\bf {x}}) =\pm \hat r$. 
If $\kappa({\bf {x}}) = \kappa (r)$ is radially symmetric, then so is
$T({\bf {x}}) = T(r)$. 
Eq.\,\ref{superfinal}
becomes
\begin{eqnarray}
\nonumber
\frac{[1 + \alpha(r)]}{2  r^{d-1}} [r^ {d-1} T'(r)]'  \hspace{2.5cm}  \\
 -  \frac{ (d-1) \alpha(r)}{ r} T'(r) 
\pm \frac{d \mu}{2 \sigma} \beta(r) T'(r)
&=& - \frac{1}{2 D}
\label{Tradial}
\end{eqnarray}
with $T(R_0) = 0$. Here the plus (minus) sign indicates motion biased towards the positive (negative) radial direction
$\hat \gamma({\bf {x}}) = \hat r$ ($\hat \gamma({\bf {x}}) = - \hat r$) and $\alpha(r), \beta(r)$ dictate the 
bias level. For all distributions listed in 
Table \ref{table1}, the limit $\kappa(r) \to 0$ 
leads to uniformly distributed reorientation angles, $\alpha(r), \beta(r) \to 0$ and Eq.\,\ref{Tradial} is solved by the purely diffusive form 
$T(r) =(R_0^2 - r^2)/ (2 d D)$. The limit $\kappa(r) \to \infty$ instead 
leads to sharply peaked reorientation
distributions,  $\alpha(r), \beta(r) \to 1$ and Eq.\,\ref{Tradial} is solved by the purely ballistic form 
$T(r \neq R_0) = (R _0 \mp r)/ \sigma$. 
For $d=2$, the MFPT under
tangential bias,  $\hat \gamma({\bf {x}}) = \hat r^{\perp}$, satisfies 
\begin{eqnarray}
\label{Ttang}
 \frac{[1 - \alpha(r)]}{2 r} [r T'(r)]'  +  \frac{\alpha(r)}{r} T'(r) = -\frac 1 {2 D}, 
\end{eqnarray}

\begin{figure*}
\centering
{\includegraphics[height=0.31\textwidth]{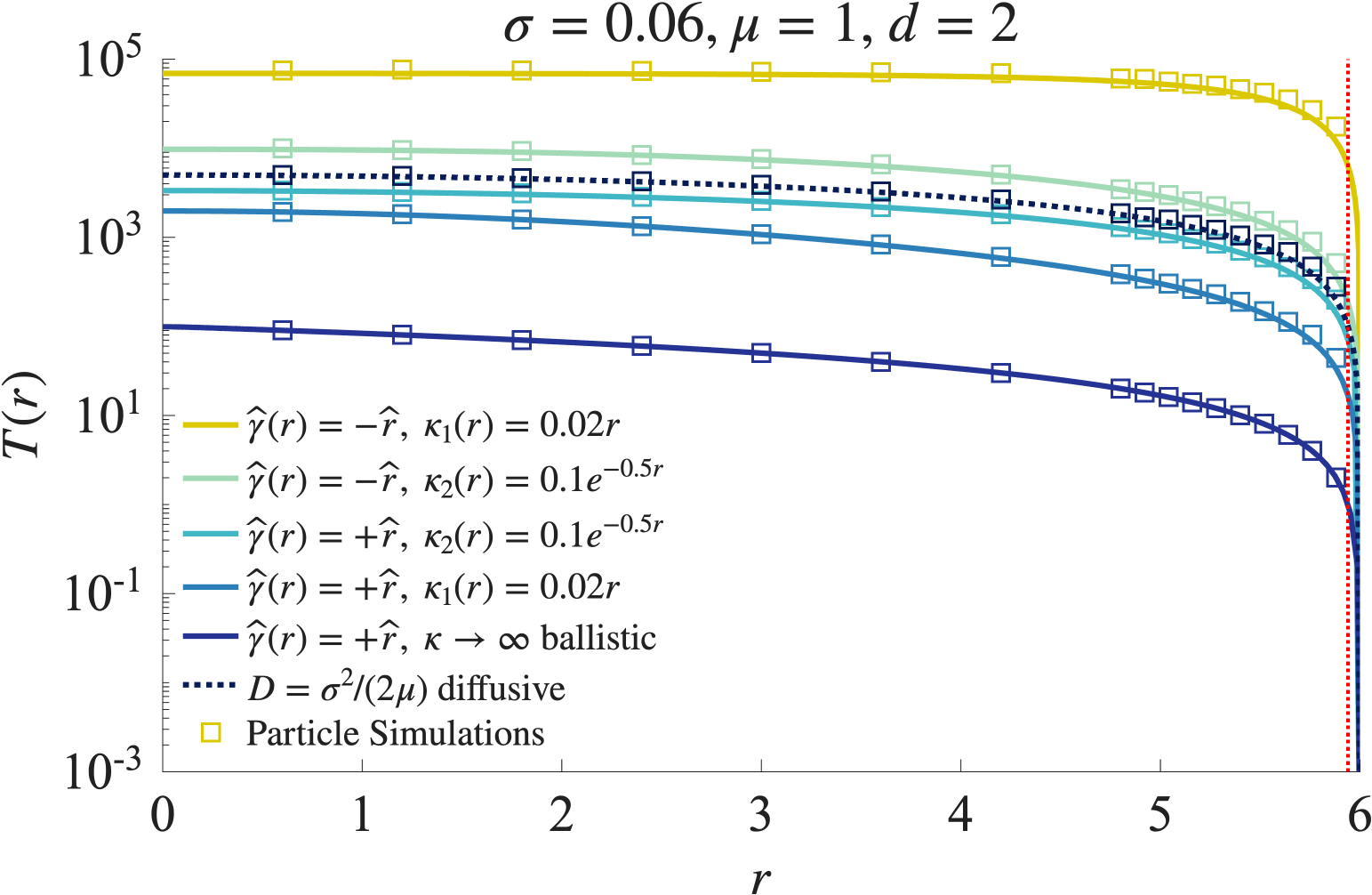}} \qquad {\includegraphics[height=0.31\textwidth]{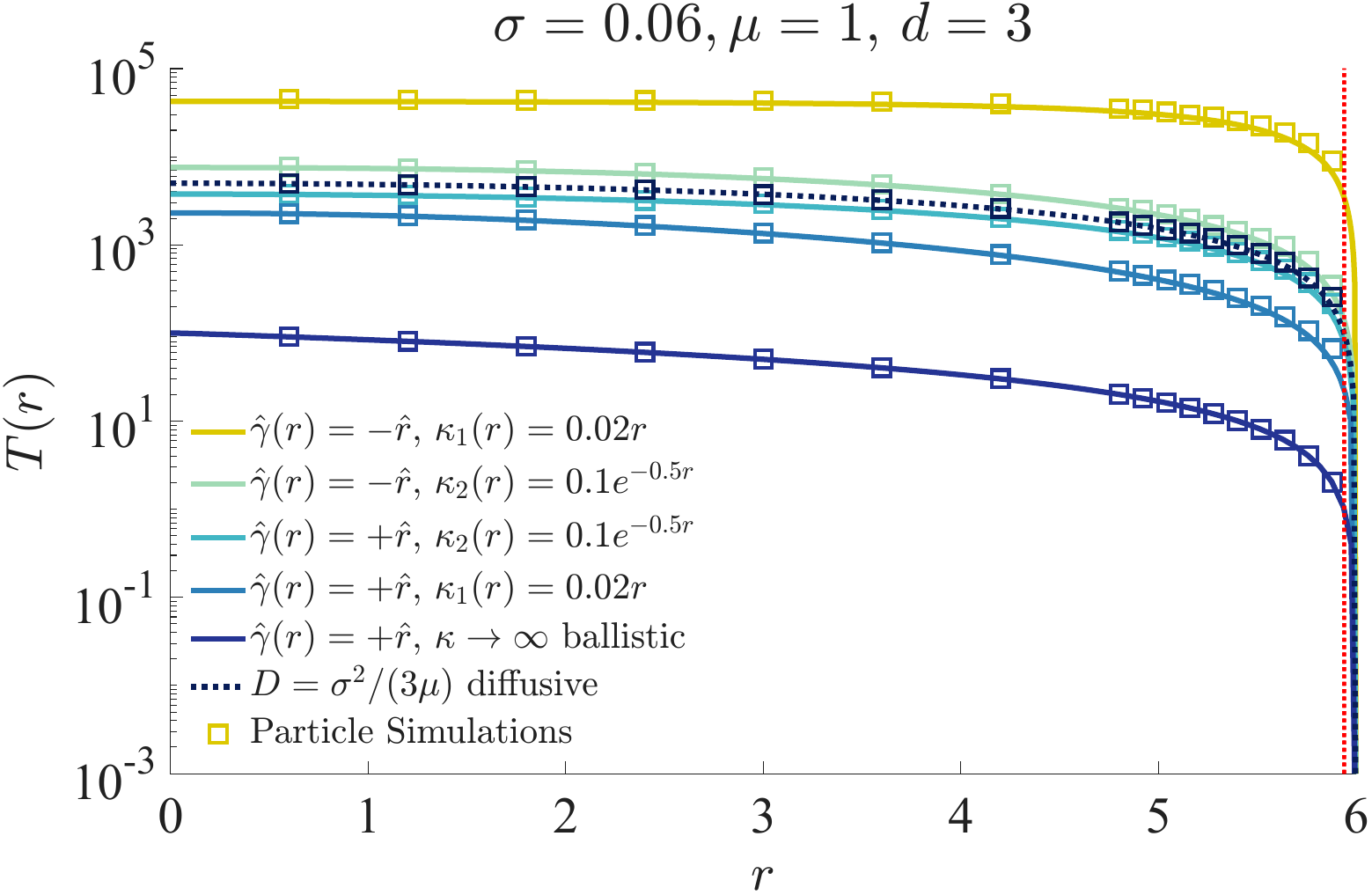}}
\caption{Analytical estimates from Eq.\,\ref{MFPTsol}
(curves) and particle simulations (squares) of the MFPT to the boundary of a ball of radius $R_0 = 6$ in $d=2$ (left) and $d=3$ (right). 
Reorientations follow the von Mises ($d=2$) or Fisher ($d=3$) distributions given in Table \ref{table1}. 
The speed $\sigma = 0.06$ and turning rate $\mu=1$ yield the diffusion constant $D = \sigma^2 / (d \mu)$, 
mean free path $\ell = \sigma/\mu = 0.06$, and Knudsen number $\varepsilon = \ell/R_0 = 0.01$. 
Deviations may arise when the initial distance to the exit boundary is comparable to $\ell$, i.e. for $r \gtrsim R_0 -\ell = 5.94$, 
denoted by the red vertical line.
For unbiased motion (black dotted curves), when $\kappa(r) = \alpha(r) = \beta(r) =0$, 
the diffusive form $T(r) = (R_0^2 - r^2)/ (2 d D)$ is recovered from
Eq.\,\ref{MFPTsol}. In all other curves 
motion is biased along the positive or negative radial direction $\hat \gamma(r) = \pm \hat r$ via
$\kappa_1(r) = r/\lambda_1$ (increasing as the boundary is approached) or $\kappa_2(r) = A e^{-r/\lambda_2}$ 
(decreasing as the boundary is approached). The limit $\kappa(r) \to \infty$, $\gamma(r) = \hat r$,  $\alpha(r), \beta(r) \to 1$ 
yields the ballistic form $T(r) = (R_0 - r)/ \sigma$.
Compared to purely diffusive trajectories, positive biases reduce the MFPT,  
negative biases induce detours that delay exit. Even relatively modest biases can alter the MFPT by several orders of magnitude. 
We set $A = 0.1, \lambda_1 =50, \lambda_2 = 2$. Simulations
are averaged over $10^4$ runs. Units are arbitrary. Other parameter choices are in the SM.}
\label{results1}
\end{figure*}

\noindent
independently of the sign of the bias.  
Eqs.\,\ref{Tradial} and \ref{Ttang}  can be solved exactly for any radially symmetric
boundary conditions. The MFPT to exit an annulus 
$(d=2)$ or a spherical shell $(d=3)$
defined by $\rho \leq r \leq R_0$ 
from either or both inner and outer boundaries is given by 
\begin{eqnarray}
T(r) &=&  H_1 +  \int^r e^ { - {\cal F} (\eta)} d \eta 
\left[ H_2  - \frac{1}{D}\int^{\eta} 
\frac {e^ {  {\cal F}(s)}} {1 + z \alpha(s)} d s \right]
\qquad
\label{MFPTsol}
\end{eqnarray}
where $z = 1$ for radial bias ($d =2,3$), 
$z = -1$ for tangential bias ($d=2$), 
the integration constants
$H_1, H_2$ depend on the chosen exit conditions, 
and 
\begin{eqnarray}
\label{MFPTIF}
{\cal F}(r) = \int^ r  \frac {(d-1)} {s} \frac {1 - z \alpha(s)}{1 +  z \alpha(s)} \pm  \frac{z+1}{2 \sigma}
\frac {d \mu \beta(s) }{1 + \alpha(s)} \ ds. \quad
\end{eqnarray}
In Fig.\,\ref{results1} we compare analytical estimates to numerical simulations of the MFPT for particles following the von Mises 
($d=2$)
or Fisher ($d=3$) distributions with preferred orientation
$\hat \gamma = \pm \hat r$ and exiting a disk
($d=2$) or sphere ($d=3$) at $r = R_0$.
This corresponds to $z=1$, $\rho=0$
in Eq.\,\ref{MFPTsol}
where $H_1, H_2$ are determined from $T'(0)=T(R_0)=0$. 
We choose $\kappa(r) \to 0$ (standard diffusion),  
$\kappa(r) \to \infty$ (ballistic motion) and 
two non-uniform concentrations
$\kappa_1(r) = r/\lambda_1$, where the bias increases 
as the boundary
is approached, and $\kappa_2(r) = A \, e^{-r/\lambda_2}$, where the opposite is true.  
If we identify $L$ with $R_0$, the parameters used in Fig.\,\ref{results1} yield $\varepsilon = \ell/R_0 =0.01$. As shown, simulations and estimates from Eq.\,\ref{MFPTsol} are in excellent agreement. 
Locally, we expect deviations only in the thin layer $R_0 - r \lesssim \ell$
when the distance to the boundary is ${\cal O}(\ell)$.
Fig.\,\ref{results1} also shows that $T(r)$ is highly sensitive to bias, as even slight directional preferences can lead to dramatic departures from the standard diffusive MFPT. 
Relative to isotropic diffusion, biases in the positive radial direction shorten the MFPT, biases in the opposite direction
delay it. 

Of particular interest is the narrow capture problem, where
a particle seeks a small target in a larger domain 
\cite{Kurella2015,CKL2025}. We assume that
the particle is initiated at $\rho < r < R_0$ and
follows the von Mises ($d=2$) or Fisher ($d=3$) distributions with uniform 
$\kappa$ (and uniform $\alpha, \beta$ as per Table\,\ref{table1}) under radial bias. We first take the bulk limit $\ell/\rho \to 0^+$, so that for 
$r \gtrsim \rho + \ell$, $T(r)$ is found by solving Eq.\,\ref{MFPTsol} with $z=1$ and
$T(\rho) =T'(R_0) =0$. We then take the narrow-capture 
limit $\rho/R_0 = \zeta \to 0^+$, effectively enlarging the bulk and further suppressing boundary effects.
The derived $T(r)$ is used to approximate the global MFPT  $\tau_{\rm g}$, 
defined as spatial average of the MFPT, 
$\tau_{\rm g} \sim |{\cal{D}}|^{-1}\int_{\cal{D}}T(r) r^{d-1} dr$, leading to the 
scaling form 
\begin{equation}
\label{scaling1}
\tau_{\rm g} \sim  C_{\rm r} \, \zeta^{\frac{d(\alpha-1)+2}{\alpha+1}}
(1 + {\cal O} (\zeta)), \quad 
\qquad\zeta\to0^{+}.
\end{equation}
This anomalous scaling differs markedly from the standard diffusive forms $\tau_{\rm g} \sim \log\zeta$ ($d=2$) and $\tau_{\rm g} \sim \zeta^{-1}$ ($d=3$); 
in particular, $\tau_{\rm g}$ remains finite as $\zeta\to0^{+}$ for all $\alpha >0$ in $d=2$ and
for $\alpha > 1/3$ in $d=3$.  The corresponding $\kappa$ can be found via Table \ref{table1}.
For tangential bias in $d=2$, a similar 
approach with $z=-1$ in Eq.\,\ref{MFPTsol} yields
\begin{equation}
\tau_{\rm g} \sim   C_{\rm t} \, \zeta^{-\frac{2 \alpha}{1- \alpha}} (1 + {\cal O} (\zeta)), \qquad\zeta\to0^{+},
\end{equation}
so that $\tau_{\rm g}$ diverges for all $0 < \alpha < 1$, another striking departure from standard diffusion. 
For $\zeta \to 0$, bulk transport dominates and  
boundary layer effects will modify only prefactors, 
not scaling exponents.
Higher moments $T_{n}({\bf x}) = n (\mu R_0^2/\sigma^2)^n \int_0 ^{\infty} t^{n-1} S_0({\bf x}, \tau) d\tau$
can also be derived from Eq.\,\ref{backwardres} to leading order in $\varepsilon$ 
via the hierarchy
\begin{eqnarray}
\label{hierarchy}
{\mathbb D}({\bf {x}}) : \nabla \otimes \nabla \,
T_n
+ {\bf {b}}({\bf {x}}) \cdot \nabla \, T_n = -n T_{n-1}
\end{eqnarray}
where $n > 1$ and $T_1({\bf x}) \equiv T(\bf x)$ is the MFPT. 
Eq.\,\ref{hierarchy} is valid for all   $d=2,3$ distributions in Table \ref{table1}.
\noindent\begin{figure*}
\centering
{\includegraphics[width=1.0\textwidth]{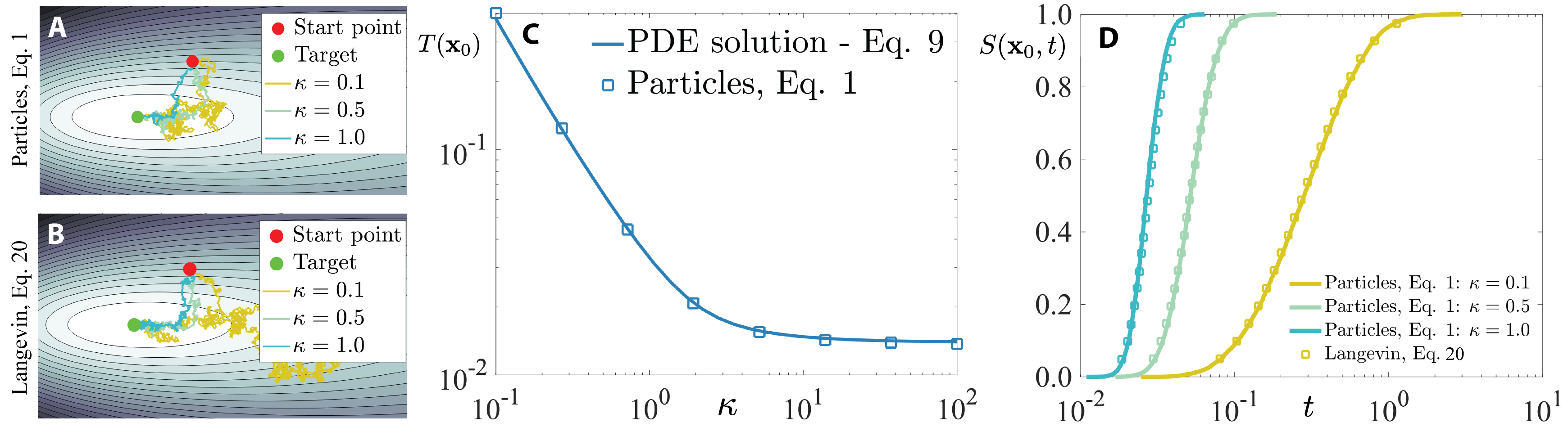}}
\caption{Two-dimensional particles searching for the locus ${\bf {x}}_T$ of maximal concentration of the
Gaussian plume $\phi{(\bf x)} = Q /(2 \pi K x) e^{-u (y^2 + H^2) /(4 K x)}$ with $Q=2$, 
 $K =0.5$, $u=5$, $H=1$ and ${\bf {x}}_T = (u H^2 / 4K, 0) = (2.5, 0)$. 
Particles start at $\bx_0=(3,0.5)$ and
follow the von Mises distribution in Eq.\,\ref{VM} with 
$\hat \gamma(\bf x) = \nabla \phi(\bf x) / | \nabla \phi(\bf x)|$ 
and uniform $\kappa$. Speed and turning rate are
$\sigma = \sqrt{\mu}= 10^2$.  The trajectories in 
panels A,B, derived from Eq.\,\ref{forward} and the Langevin process in Eq.\,\ref{Langevin} 
respectively, are qualitatively similar. Panel C shows strong agreement between the MFPT $T({\bf x}_0)$ from Eq.\,\ref{superfinal} and simulations of Eq.\,\ref{forward} averaged over $10^4$ runs. Panel D shows that Eq.\,\ref{Langevin}
accurately captures the survival probability $S({\bf x}_0, t)$.
Units are arbitrary.
}
\label{results2}
\end{figure*}

\noindent
Upon expressing Eq.\,\ref{superfinal} in operator form, 
${\cal L^*} T{(\bf x)} = -1$, and using  Itô's interpretation, we infer the following
Langevin process 
%
\begin{equation}\label{Langevin}
d\bx = { \bf b}(\bx)dt + [2\mathbb{D}(\bx)]^{\frac12}dW_t, 
\end{equation}
where ${\bf b(\bx)}, \mathbb{D}(\bx)$ are as in 
Eqs.\,\ref{diffVM}, \ref{driftVM}, 
the root is taken in the matrix sense, and $W_t$ is a vector of
Brownian motions.  
For $\varepsilon \ll 1$ and by construction, 
the velocity jump and Langevin processes 
display identical first-passage statistics;  Eq.\,\ref{Langevin} 
allows for the exploration of additional dynamical properties.
As an example, we consider 
a random walker 
seeking the locus ${\bf x}_{\rm T}$ of maximal concentration of a given plume $\phi(\bf x)$ in $d=2$. 
We let the walker follow the von Mises distribution in Eq.\,\ref{VM} with uniform $\kappa$ and 
preferred direction given by the normalized gradient 
of $\phi({\bf x})$,  $\hat \gamma(\bf x) =  \nabla \phi(\bf x) /  |\nabla \phi(\bf x) | $.  
For concreteness, 
we choose $\phi({\bf x})$ as the planar projection at steady state of a 
Gaussian plume generated at altitude $H$, at constant rate $Q$, advected at speed $u$ along the $x$-direction, and dispersed with diffusion constant $K$, leading to $\phi{(\bf x)} = Q / (2 \pi K x) e^{- u (y^2 + H^2) / (4 K x)}$ \cite{Stokie2011}; 
this plume is maximal at ${\bf{x}}_{\rm T} = (u H^2 / 4 K,0)$.
Representative trajectories from simulations of particles starting at 
${\bf x}_{0}$ following Eq.\,\ref{forward} or the Langevin dynamics in Eq.\,\ref{Langevin} are shown in Figs.\,\ref{results2}A and \ref{results2}B, respectively. The corresponding MFPTs to ${\bf x}_{\rm T}$, calculated from 
Eq.\,\ref{forward}, 
are displayed in Fig.\,\ref{results2}C as a function of $\kappa$ and
show excellent agreement with solutions to Eq.\,\ref{superfinal}.  
As seen in Fig.~\ref{results2}D, 
the Langevin process in Eq.\,\ref{Langevin} yields the full survival probability $S({{\bf {x}}_0}, t)$,
and not just the MFPT $T({\bf x}_0)$, with remarkable accuracy.

Our results have broad applications. As shown in the SM, the von Mises kernel in Eq.\,\ref{VM} 
is the steady-state distribution of an overdamped angular system driven by the 
potential $U(\theta, {\bf x}) = - \tau_{\rm M}({\bf x})  \cos (\theta - \gamma({\bf x}))$, 
where the torque biases orientation along $\hat \gamma({\bf x})$ in competition with 
rotational noise \cite{Hofling2025}, leading to $\kappa({\bf x}) \propto \tau_{\rm M}({\bf x})$. Assuming an electric field of magnitude $E({\bf x})$ directed along $\hat \gamma({\bf x})$, we can model $\tau_{\rm M}({\bf x}) \propto E({\bf x})$, 
so that $\kappa({\bf x}) \propto E({\bf x})$. 
Thus, Eq.\,\ref{VM} arises as a natural way to describe field-driven motion.
For $d=2$, radial guidance and a uniform, weak $E$
Eq.\,\ref{scaling1} yields $\tau_{\rm g} \sim \zeta^{\nu}$ with $\nu \propto E^2$, providing an avenue to 
test the predicted scaling. For behavioral systems, without physical torques, Eq.\,\ref{VM} can still be used to
model transport biased by trails, social cues, or learning.  
From this perspective, as discussed in the SM, our work offers a unified framework 
in which to study the MFPT of field-driven colloidal matter,  run-and-tumble bacteria under chemotaxis, active swimmers, and animal movement
\cite{Saito2025, Karani2019, Codutti2019, Popescu2018, Buness2024, Sourjik2012, Mlynarczyk2019, Nieuwenhuizen2004, McKenzie2009, Boissard2013, Lauga2006, Kummel2013, Schneirla1944, Narazaki2021, Wang2013}. Finally, mapping the dynamics onto Eq.\,\ref{Langevin} 
allows to reconstruct the bias terms $\alpha({\bf{x}})$  and 
$\beta({\bf x})$ from experimentally observed trajectories.

In summary, we presented a novel theory for the MFPT of velocity jump processes
in higher dimensions, whose general form is
Eq.\,\ref{MFPT}.  For fixed speed and the angular distributions
listed in Table \ref{table1} for $d=2,3$, we show that the capture statistics in the small Knudsen number limit are governed by the bias terms $\alpha(\bx)$ and $\beta(\bx)$ via Eq.\,\ref{superfinal}, and 
can differ quite significantly 
from those observed under standard diffusion. Furthermore, the identification of a generator function 
for the underlying stochastic process leads to the Langevin approximation in Eq.\,\ref{Langevin} 
that efficiently describes the transport dynamics. 
We fixed particle speed and let directional persistence arise from the sharpness of the angular distribution. If particle speeds followed heavy-tailed distributions, as in L\'evy processes, exceptionally large jumps could lead to rare boundary-reaching events. This scenario would require a different analysis than what presented here, invoking for example the “single big jump” principle \cite{Vezzani2019}. Solving the full problem in Eq.\,\ref{backward} requires tracking both particle position and velocity. Appropriate boundary conditions must be specified for $S({\bf x}, {\bf v}, t)$, carefully distinguishing between inward and outward directions on the boundary, as shown in the SM. A full treatment is left for future work.

\begin{acknowledgments}
We thank T. Chou and K. Painter for fruitful discussions. 
This work was supported by the ARO (grant W911NF-23-1-0129, MRD), the NSF (grants 
DMS-2052636, AEL and OAC-2320846, MRD) and the NSERC 
(Discovery grant RGPIN-2023-04269, TH).
The data that support the findings of this article are openly available \cite{GH2026}. 

\end{acknowledgments}








\end{document}